# A Lip Vibration Model Using Mechanical Properties of Flesh

Michael T. Strauss[1,2,a]

[1]HME Newburyport, MA, 01950 USA

[2]Phillips Exeter Academy, Exeter, NH 03833 USA

[a] email: mstrauss@alum.mit.edu

## Abstract

The mechanical properties of the human lip between relaxed and fully contracted are not enough to explain the full range of professional brass players.   Brass players manipulate their mouth cavity and vocal tract which act as Helmholtz resonators driving the lips.   Brass players increase their upper ranges by reducing the amount of lip mass that vibrates using upstream or downstream techniques.   In addition, it was found that the cup of the mouthpiece is a separate resonating cavity: not just an extension of the instrument funneling air down the length of tube.  For the lowest three octaves of the range, the trombone does not act as a tube with a standing wave resonance, but as a transmission line responding to a string of pulses produced by the lips.

## Materials and Methods

For pressure measurements, a Bach 36 trombone with closed wrap F attachment and a Shilke 50 mouthpiece (25.4 mm diameter cup) were used.   Holes were drilled into the cup and backbore of the mouthpiece and PiezoBarrel® "Wood" pickups (PiezoBarrel, Brisbane Australia) were mounted, in both.  1.5" (0.3825 M) of ¼" ( 0.006375) outer diameter, 3/16" (0.004781 M) inner diameter plastic tubing was attached to a third PiezoBarrel pickup.  The other end of the tube was inserted into the mouth of the trombone player.  An Audio-Technica PRO35 clip-on microphone was attached to the bell of the trombone.  The three pickups and microphone were plugged into a Focusrite Scarlett 18i8, which was attached to an Apple Mac laptop with a USB cable.  The signals were captured simultaneously and analyzed with Audacity, a free, open source, cross-platform audio software product.  To measure the popping frequency of the mouthpiece, it was slapped against the palm of the hand and the decaying sinusoid was measured with the fastened pickups, also with Audacity.

## Introduction

Musician's lips are the vibration source for brass instruments.  Slow-motion photography shows lips vibrating at the same frequency as the sound[1,2] and brass players can produce lip vibrations without the instrument or mouthpiece.  Lips are often described as a type of resonating mass and spring system[3,4,5].  From the scientific side, models include one mass model, sliding door, swinging door, inward and outward striking reeds, and spring-loaded pressure release valves[3,4,5].  However, none of these models use the intrinsic mechanical properties of the lips.  They all assume that the lips are vibrating at some frequency and there must be some combination of mass and spring stiffness, that can produce it, without ever specifying values.  A recent text on the acoustics of musical instruments summarizes the state of lip vibration models[6]: "We conclude that it is difficult to develop an accurate and reliable model."  From the brass instrument pedagogy, teachers can tell students what they need to do unrelated to physical principals behind them.  The work here attempts to address these problems by creating a model of lip vibration based on material properties of flesh combined with manipulations of the mouth, tongue, lips, and vocal tract that brass players use to produce sound.

Lips are not simple mass and spring systems. An explanation of how lips vibrate should explain the full range brass musicians play. For the simplest continuum mass and spring system, the resonance frequency is proportional to $\sqrt{E/\rho}$ where $E$ is Young's modulus and $\rho$ is the density in mass per volume.

The density of a lip is essentially the same for uncontracted and contracted muscles. One can define the range as the ratio of the highest note a player can produce to the lowest, giving:

$$Range = \sqrt{E_{Highest}/E_{Lowest}} \tag{1}$$

Modulus measurements of uncontracted muscle tissue are generally 10-50 kPa[7]. Ultrasonic measurements of the lower lip average 33.7 kPa[8]. Measurements of fully contracted muscle are rare. Measurements on human biceps show a linear relationship between shear modulus and maximal voluntary contraction (MVC) up to 30% of MVC[9]. Extrapolating to 100% contraction, the ratio is 90, giving a range of 33.7-303 kPa for uncontracted and contracted lip tissue. Similar measurements on the little finger for the full range from 0-100% of MVC show linearity in the entire range[10], but a lower modulus ratio of 25. For isotropic materials, the shear modulus and Young's modulus are proportional. The ratio of the highest to lowest playable frequencies would be 5 for the little finger and approximately 9.5 for biceps. Assuming the lips fall somewhere between the two, that gives a playable range of a little more than 2 octaves to a bit more than 3 octaves, because each octave represents a factor of 2 in frequency.

A professional brass player may have a five-octave range or more[11]. Therefore, this simple model fails. A brass player would have to be able to contract their lip muscles to a modulus 16 times greater than their biceps just to explain a five-octave range. This is unlikely.

### Farkas-Arban-Leno Model

Following the descriptions in brass instruction books, the lips are modeled as two flat semi-circular vibrating plates. In "The Art of Brass Playing[12]," Farkas states "We are concerned with having the upper and lower lips directly opposite each other, in an up and down consideration so that they abut together, without one lip (usually the lower) sliding behind the other. The foundation or support of the lips is the responsibility of the upper and lower front teeth. Therefore, if the lips are to line up so that they abut together without sliding one under the other, the upper and lower front teeth must also be exactly in line… the mouthpiece assumes a fairly horizontal position… so that the mouthpiece pressure is exactly and evenly distributed on both lips." It is interesting that while stating this method of playing, none of the photographs of himself or the other brass players of the Chicago Symphony included in the book support it. However, this suggests each lip is a uniform semi- circular plate, the dimensions are set by the mouthpiece inner diameter and lip thickness.

For boundary conditions, several are suggested by the Arban's method[13]. Originally for trumpet and cornet, it has been transcribed for trombone, tuba, French horn and translated into many languages. It is a widely used brass instruction and exercise text, has been in print for over 150 years, with new editions produced periodically. Practically every brass player has used it at some point in their development. Arban writes, "In order to produce the higher notes, it is necessary to press the instrument against the lips, so as to produce an amount of pressure proportionate to the needs of the notes to be produced; the lips being thus stretched, the vibrations are shorter, and the sounds are consequently of a higher nature." While this advice of pressing the mouthpiece against the lips harder to produce higher notes remains in many modern versions of the text, it is not recommended today [14]. Using too much pressure is a common problem for many players, particularly beginners. However, it suggests clamped boundary conditions around the circumference at high frequencies.

Similarly, for playing in the low register Arban states "For descending passages it is necessary to apply the mouthpiece more lightly, in order to allow a larger opening for the passage of air. The vibrations then become slower owing to the relaxation of the muscles, and lower sounds are obtained in proportion to the extent to which the lips are opened." Therefore, for low register playing, the mouthpiece is lightly resting on the lips, suggesting simply supported boundary conditions. In summary, Arban recommends contracting the lips and pushing the mouthpiece into them for high notes and relaxing the lip muscles and easing off the pressure for low notes. He states doing this, a trumpet player can reach a high C (C5 in scientific pitch notation) but he recommends not playing in that register too much as it can tire and damage the lips. Therefore, in the high range, the lips have their highest modulus and a clamped boundary condition, while in the low register, they have their lowest modulus and simply supported boundary conditions.

Leno used high speed photography through clear mouthpieces showing brass players' lips have no nodal lines inside the rim of the mouthpiece when playing[1,2] meaning the lips move in the lowest mode.

In summary the model is: The lips are sandwiched evenly between the teeth and the rim of the mouthpiece. The air pressure from the lungs pushes on the lips and starts an oscillation. The lips are modeled as two semi-circular flat plates and the question is to find the resonant frequency as a function of the Young's modulus, density, geometry, and boundary conditions. Arban's method suggests pressing the mouthpiece against the lips in the high register which is represented by clamped boundary conditions where both the displacement and slope of the displacement are zero at the mouthpiece. For low notes, Arban suggests the mouthpiece rests lightly on uncontracted lips. This is represented by simply supported boundary conditions, where there is no displacement at the boundary, and the bending moment is zero. The diameter separating the lips is the vibrating surface which is a free boundary condition. Using Leno's observation, choose the lowest mode of vibration. Following the established analysis for circular flat plates, the equation governing the normal deflection, $W$, of flat plates is[15]:

$$D\nabla^4 W(r,\theta,t) + \rho_A \frac{\partial^2 W(r,\theta,t)}{\partial t^2} = 0 \qquad (2)$$

With:

$$D = \frac{Eh^3}{12(1-\nu^2)} = \text{ the flexural rigidity} \qquad (3)$$

And $\rho_A = \rho h$ is the density, mass per area, $\rho$ is the density, mass per volume, $E$ is Young's modulus, $h$ is the thickness of the lips, and $\nu$ is poison's ratio.

Assuming sinusoidal time dependence, $W(r,\theta,t) = w(r,\theta)e^{i\omega t}$, with a frequency $\omega$, the equation becomes:

$$\nabla^4 w(r,\theta) - k^4 w(r,\theta) = 0 \tag{4}$$

With:

$$k^4 = \frac{\rho_A \omega^2}{D} = \frac{12\rho\omega^2(1-\nu^2)}{Eh^2} \tag{5}$$

$$\omega = k^2 \sqrt{\frac{D}{\rho_A}} = \frac{k^2 h}{2}\sqrt{\frac{E}{3\rho(1-\nu^2)}} \tag{6}$$

Factoring the equation gives:

$$(\nabla^2 + k^2)(\nabla^2 - k^2)w(r,\theta) = 0 \tag{7}$$

Which is equivalent to:

$$(\nabla^2 + k^2)w(r,\theta) = 0, \text{ or} \tag{8a}$$

$$(\nabla^2 - k^2)w(r,\theta) = 0 \tag{8b}$$

The general solution in polar coordinates is:

$$w(r,\theta) = \sum_{n=0}^{\infty}[a_n J_n(kr) + b_n Y_n(kr) + c_n I_n(kr) + d_n K_n(kr)]\cos(n\theta)$$

$$+ \sum_{n=1}^{\infty}[e_n J_n(kr) + f_n Y_n(kr) + g_n I_n(kr) + h_n K_n(kr)]\sin(n\theta)$$

$$\tag{9}$$

Where $J$ and $Y$ are Bessel functions of the first and second kind respectively and $I$ and $K$ are modified Bessel functions of the first and second kind respectively and $a, b, c, d, e, f, g,$ and $h,$ are constants. The solution must be finite at $r = 0$, therefore, the coefficients to $Y$ and $K$ are zero. Choosing the lowest mode, $n = 0$, the solution simplifies to:

$$w(r) = a_0 J_0(kr) + c_0 I_0(kr) \tag{10}$$

The free boundary conditions along the diameter are the angular component of the bending moment ($M_\theta$) and the angular component of the Kelvin-Kirchoff edge reaction ($V_\theta$) are both zero. In terms of the transverse shearing force ($Q_\theta$) they are:

$$M_\theta = -D\left[\frac{1}{r}\frac{\partial w}{\partial r} + \frac{1}{r^2}\frac{\partial^2 w}{\partial \theta^2} + \nu\frac{\partial^2 w}{\partial r^2}\right] = 0 \tag{11}$$

$$V_\theta = Q_\theta + \frac{\partial M_{r\theta}}{\partial r} = 0 \tag{12}$$

$$M_{r\theta} = -D(1-\nu)\frac{\partial}{\partial r}\left(\frac{1}{r}\frac{\partial w}{\partial \theta}\right) \tag{13}$$

$$Q_\theta = -D\frac{1}{r}\frac{\partial}{\partial \theta}(\nabla^2 w) \tag{14}$$

The edge reaction will be identically zero because $Q_\theta$ and $M_{r\theta}$ have no angular dependance. However, the bending moment will not necessarily be zero. In a similar situation with a single free surface on a rectangular plate the bending moment requirement is discarded because it would give no displacement along the free surface[16].

Around the inner circumference at the mouthpiece, $r = R$, the displacement is zero. Using a clamped boundary condition, the derivative is also zero:

$$w(R) = 0 = a_0 J_0(kR) + c_0 I_0(kR) \tag{15}$$

$$\frac{\partial w}{\partial r}(R) = 0 = a_0 \frac{\partial J_0}{\partial r}(kR) + c_0 \frac{\partial I_0}{\partial r}(kR) \tag{16}$$

With $\lambda = kR$, and the relationships for derivatives of Bessel functions[17], the conditions become:

$$\begin{bmatrix} J_0(\lambda) & I_0(\lambda) \\ -\lambda J_1(\lambda) & \lambda I_1(\lambda) \end{bmatrix} \begin{bmatrix} a_0 \\ c_0 \end{bmatrix} = 0 \tag{17}$$

This is satisfied if the determinant of the matrix is zero, giving the characteristic equation:

$$J_0(\lambda) I_1(\lambda) + J_1(\lambda) I_0(\lambda) = 0 \tag{18}$$

Numerically, the lowest positive value solution is the fundamental frequency and is:

$$\lambda_{Clamped} = 3.196, \qquad \lambda^2_{Clamped} = 10.22 \tag{19}$$

Using a simply supported boundary condition, the displacement and the radial component of the bending moment are zero at the inner circumference of the mouthpiece:

$$w(R) = 0 = a_0 J_0(\lambda) + c_0 I_0(\lambda) \tag{20}$$

$$M_r(R) = 0 = -D \left[ \frac{\partial^2 w}{\partial r^2} + \nu \left( \frac{1}{r} \frac{\partial w}{\partial r} + \frac{1}{r^2} \frac{\partial^2 w}{\partial \theta^2} \right) \right]_R = a_0 \left[ J_0''(\lambda) + \frac{\nu}{\lambda} J_0'(\lambda) \right] + c_0 \left[ I_0''(\lambda) + \frac{\nu}{\lambda} I_0'(\lambda) \right] \tag{21}$$

Again, using the relationships for derivatives of Bessel functions, the conditions become:

$$\begin{bmatrix} J_0(\lambda) & I_0(\lambda) \\ -\lambda J_0(\lambda) + (1-\nu) J_1(\lambda) & \lambda I_0(\lambda) - (1-\nu) I_1(\lambda) \end{bmatrix} \begin{bmatrix} a_0 \\ c_0 \end{bmatrix} = 0 \tag{22}$$

In this case, the determinant is explicitly a function of Poisson's ratio for the lips, which have been measured to be 0.49[18]. Numerically, the lowest positive value solution is:

$$\lambda_{Simply\ Supported} = 2.280, \ \lambda^2_{Simply\ Supported} = 5.2 \tag{23}$$

The resonance frequency $\omega_0$ or $f_0$, in cycles per second, may be rewritten in terms of the mouthpiece diameter, $d$, and lip thickness, $h$, as:

$$f_0 = \frac{\omega_0}{2\pi} = \frac{\lambda^2 h}{\pi d^2} \sqrt{\frac{E}{3\rho(1-\nu^2)}} \tag{24}$$

The value of $\lambda^2$ changes by a factor of 1.96 or a little less than of 2 between simply supported and clamped boundary conditions. Therefore, pushing the mouthpiece into the lips, while keeping everything else constant increases the frequency by approximately one octave, but that is all. Pushing will not explain the ability of brass players to have a five or six octave range. This one octave increase was confirmed by the author, a professional trombone player, by buzzing a Bb3 into a mouthpiece and then pushing the mouthpiece into his lips while keeping the lip tension and air pressure constant. The frequency went up by an interval of a ninth, or a factor of 2.12.

Figure 1 shows the frequency as a function of mouthpiece diameter for this model. Typical mouthpiece diameters[19] are labeled for trumpet and French horn (16mm), trombone (25.4 mm) and tuba (32 mm). There are two curves representing the highest and lowest playable notes. For the upper limit, a Young's modulus of 303 kPa representing contracted bicep muscles and clamped boundary conditions were used. For the lower limit, a Young's modulus of 33.7 kPa representing uncontracted lip muscles and simply supported boundary conditions were used. In all cases a lip thickness of 8mm[20,21], and Poisson's ratio of 0.49 were used. The model accurately predicts that the trumpet is about an octave higher than the trombone which is an octave higher than the tuba. All the instruments have approximately four octave ranges, beginning at their respective double pedal tones, that is an octave lower than the instrument's fundamental. The model predicts the upper limit of the trumpet as a high C, which is consistent with Arban's range limit, but lower than what brass players routinely play[22].

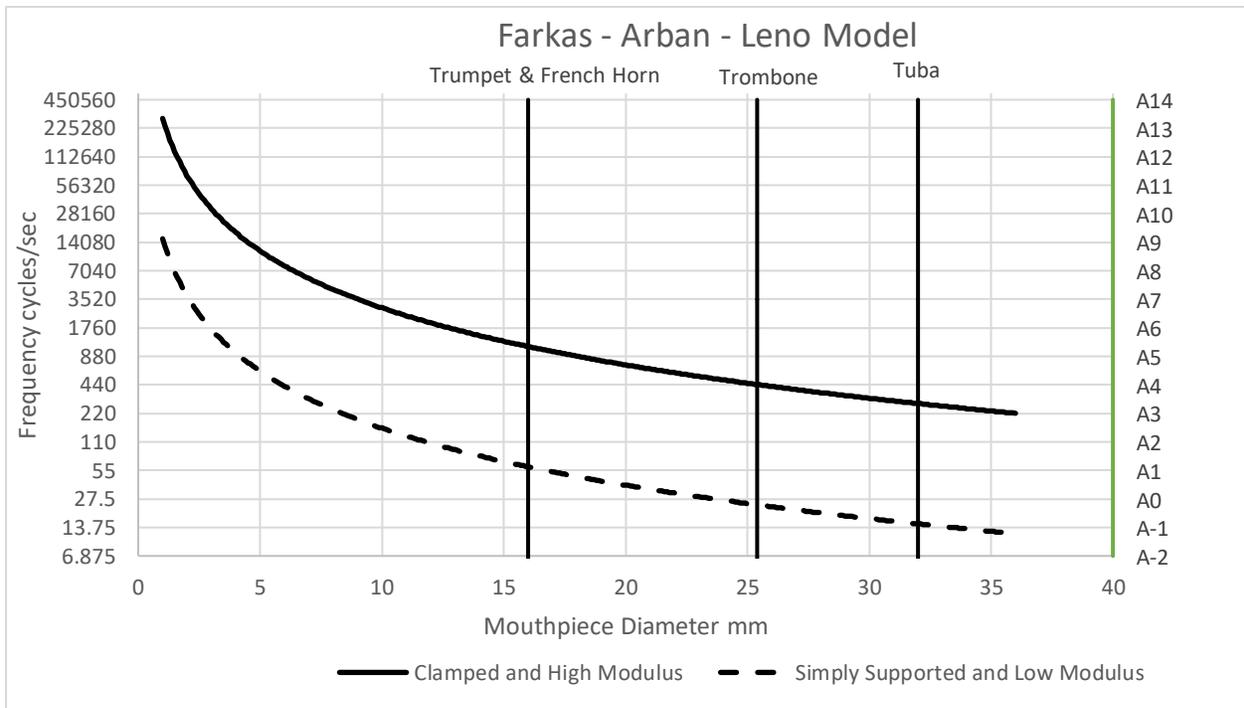

*Figure 1.* The resonance frequency (cps) versus mouthpiece diameter. Typical mouthpiece diameters are shown for trumpet, French horn, trombone, and tuba. The lower curve is for a simply supported boundary condition using Young's modulus for uncontracted muscle and the upper curve is for a clamped boundary condition using the Young's modulus for fully contracted muscle.

## Upstream and Downstream Playing

A simple mass and spring model of the lips predicts a maximum of a 3-octave range for a brass player. The Farkas- Arban-Leno model increases it to a 4-octave range, but still not enough to explain the full range of brass playing. Both models require musicians to tighten their lips to the maximum to play the highest notes. While possible, contracting muscles to their maximum is fatiguing and brass musicians would not be able to play for extended periods. Another option is to modify the amount of mass that vibrates. Experienced brass players reduce the amount of their lips that vibrate inside the mouthpiece as they move to higher frequencies[11,12,22,23]. This allows them to increase their upper range without muscle strain.  This is like how vocalists reduce the amount of the vocal folds that vibrate to reach higher notes[24]

Reinhardt[23] observed nine different embouchure types, but they all can be classified into either upstream or downstream, which are mirror images of one another.  In downstream playing, the most prevalent style, as the player moves to higher frequencies, the lower lip curls over the lower teeth and the upper lip moves over and in front of the lower lip.  This stops the lower lip from vibrating and shortens the aperture reducing the amount of upper lip vibration.  It is called downstream because the airstream is pointed downwards instead of horizontally.  In upstream playing, as the player moves to higher frequencies, the upper lip curls over the upper teeth and the lower lip moves over and in front of the upper lip.  This stops the upper lip from vibrating and shortens the aperture reducing the amount of lower lip vibration.  The airstream is directed upwards.  In downstream playing, the upper lip is the dominant lip and in upstream playing, the lower lip is the dominant lip, that is, the primary vibrating lip.

Leno[1,2] observes these two styles of playing in slow motion studies of lip vibration.  In both cases, the players push their lips toward each other.  In addition, he observed the lips are usually off center with more of the dominant lip in the mouthpiece. Both authors note the direction of the airstream has changed and less lip is vibrating.  However, neither Rheinhardt nor Leno mention that this technique is what allows players to reduce the vibrating mass, their focus is on the direction of the airstream.

There are four things happening to reduce the amount of lip that vibrates:

1. The free surface between the lips is not along the diameter of the mouthpiece, shortening the vibrating surface by reducing the length of the aperture.
2. One lip is clamped by the teeth, vibrating less or not at all, reducing the aperture height and air flow.
3. The dominant lip near the edge of the mouthpiece is pressed into the clamped lip, restricting its vibration by reducing the length of the aperture.
4. The dominant lip overlaps the clamped lip.  The center of the dominant lip is not compressed and is still free to vibrate.

One way of viewing this is to think of a shorter length aperture as a smaller mouthpiece.  The player simulates a smaller mouthpiece by clamping parts of the lip, so they don't vibrate.  Just tightening the lip muscle (orbicularis oris) won't do this because it only changes the modulus.   Pressing the lips together on their own will not increase the frequency because in-plane compressive stresses decrease the vibrational frequencies of plates[15,25].  Therefore, players resort to up and down stream techniques, using other muscles such as the metalis and depressor septi nasi to push the lips together without straining. The aperture is shortened two ways, numbers 1 and 3 above.  These can be combined to get a total shortening of the aperture, giving an effective smaller mouthpiece diameter.

Using the terminology of brass players, $f$ is the fraction of the diameter that the separation is shifted.  For example, 2/3 upper lip in the mouthpiece, $f$ = 2/3.  When the separation between the lips shifts off the center line, the length of the lip separation can be calculated as in Figure *2*[26].

$$x + R = (2R)f \qquad (25a)$$

$$x = R(2f-1) \qquad (25b)$$

From the figure, L can be found from the triangle and the ratio *L/R* is equal to the ratio *2L/D*, where *D* is the diameter of the mouthpiece:

$$L = \sqrt{R^2 - x^2} = 2R\sqrt{f - f^2} \quad (26a)$$

$$L/R = 2\sqrt{f - f^2} \quad (26b)$$

According to Leno[1,2], the players he observed had between 2/3 and 3/4 of their dominant lip in the mouthpiece (*f = 0.66 – 0.75*) when playing a high B flat (466 cps). Using the above equation, the length of the aperture is reduced between 6% and 14%, a modest reduction. He also noted that the ratio of the aperture lengths when playing a Bb4 (466 cps) to Bb3 (233 cps)) was between 1:1.1 and 1:1.8 for various players due to clamping. This corresponds to an aperture reduction between 10%-44%. Combining these two effects, the aperture was reduced between 14% and 52%. Therefore, notes which would normally require the lips to be 100% contracted, can be played with less muscle strain using upstream or downstream techniques. For example, referring to Figure 1, reducing an aperture to half the mouthpiece diameter would allow the highest playable note to go up by two octaves. If the aperture diameter is reduced by a factor of 4, a note that previously required clamping the lip with the mouthpiece and straining muscle to its fullest extent, can be played with no pressure and a relaxed lip. Experienced brass players use upstream and downstream playing to drastically reduce the length of the aperture, allowing them to play exceedingly high without mouthpiece pressure or high muscle strain. In fact, some virtuosic players use the term "pinhole" to describe the aperture size for extremely high notes[11].

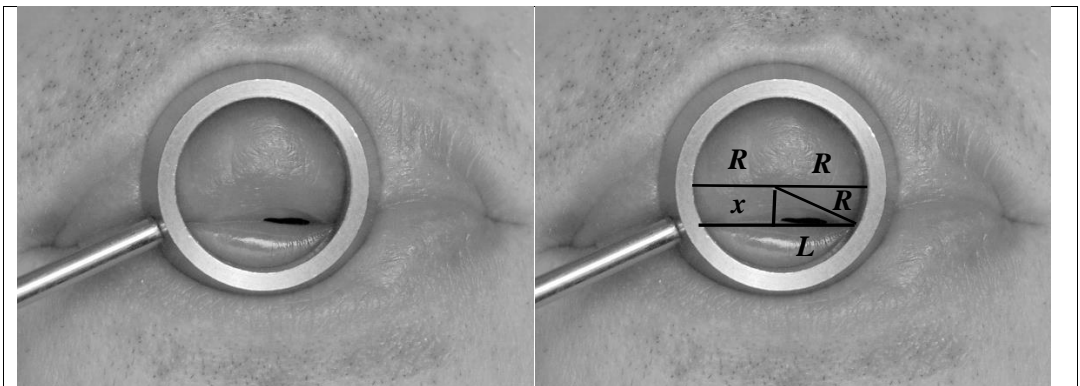

Figure 2 Moving the lips off center, reduces the length of the aperture. Permission granted J. Ericson

### Damping and Driven Oscillation

While the Arban-Farkas-Leno model modified for up and downstream playing accurately describes the range of a brass player, it is missing an important mechanism: damping. In a review of mechanical properties of lips[27], it has been found that measured value of the Quality factor for flesh ranges from 0.46 to 1.8, while for artificial lips, typically water filled latex tubes, it ranges from 3-10. Quality factors of 0.5 or less are considered overdamped[28], even with values up to 2, oscillators are very close to overdamped as shown in Figure 3. While oscillators with Quality factors in the range measured for artificial lips are clearly underdamped. Simple observations confirm that lips are overdamped. For example, one may pull either the lower or upper lip out, let go, and let it return to its resting position. The lips do not oscillate back and forth. They just go back into place and stop, a sign of overdamping[28]. This means the lips must be driven to vibrate, they will not vibrate freely on their own. The preceding resonance frequency analysis is still relevant because power is maximally delivered to overdamped oscillators at the resonance frequency they would have without any damping[28]. Therefore experienced players are able to match the resonnance of their lips to the resonance of the driver.

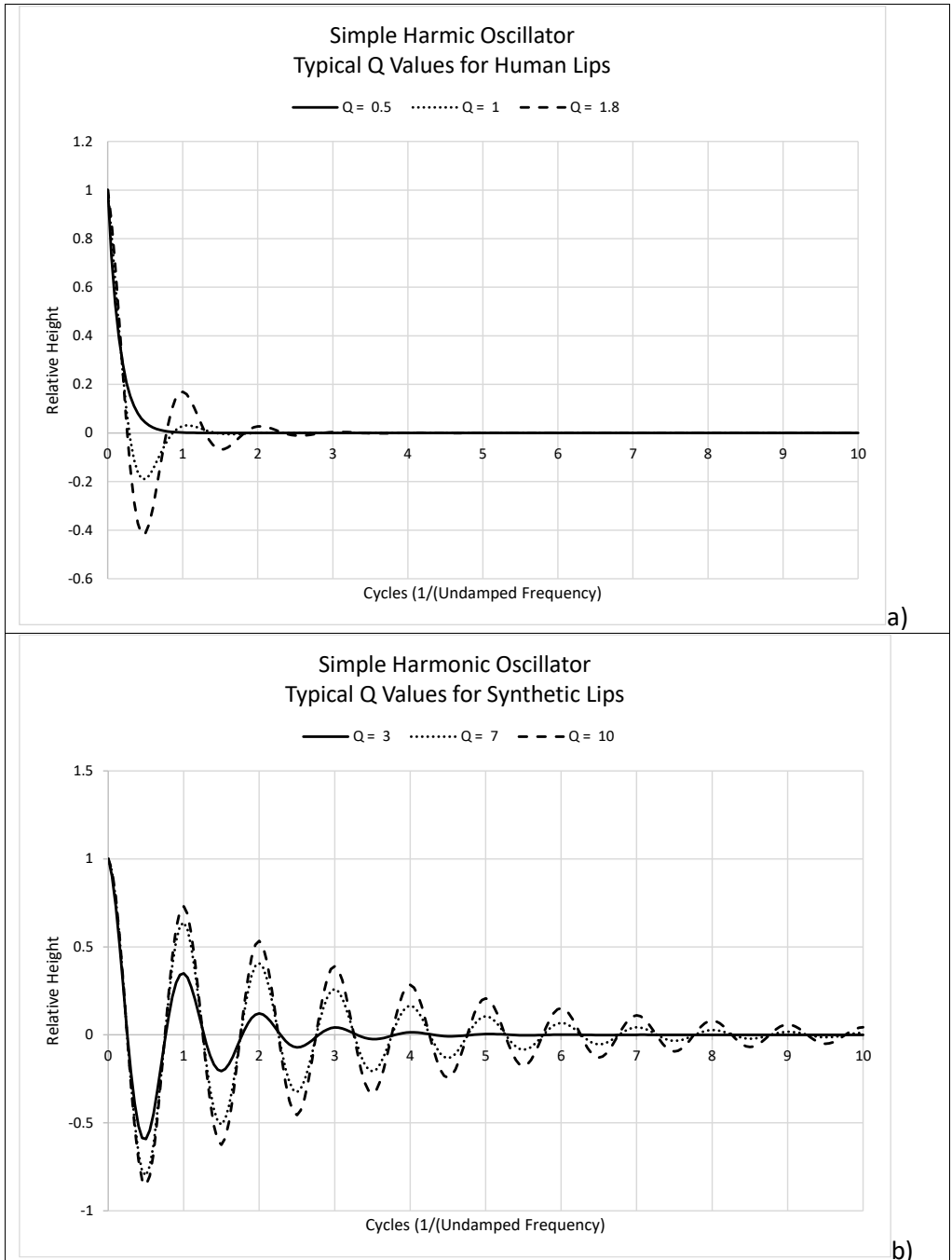

*Figure 3.* a) is a graph of a simple harmonic oscillators with the range of Q factors measured for flesh. All are or are very close to being overdamped. b) is a graph of a simple harmonic oscillator with the range of Q factors measured for artificial lips and they are underdamped.

It has been hypothesized that lips are like spring loaded pressure relief valves: the player's lung pressure forces the lips open and Bernouli's principle reduces the pressure at the aperture which closes them[3,5,29]. However, for Bernouli's principle to be applicable, the fluid flow must be inviscid, laminar, and in steady state[30]. Air is a good example of an inviscid fluid. At the aperture, no measurements have been done to show the fluid is laminar and not turbulent. Finally, the flow is definitely not in steady state because the lips are openning and closing fifty to several thousand times per second. Therefore the air stream at the lips is stopping and starting fifty to several thousand times per second, right where one would like to apply Bernouli's principle. The flow is not in steady state, therefore there are no streamlines, and Bernouli's principle is inapplicable. In fact, using Bernouli's principle will get the phase wrong between the pressure and velocity. In sound waves pressure and velocity are 90° out of phase because the pressure and velocity are derivatives of one another[31]. However Bernouli's principle states the sum of the pressure and the kinetic energy (square of the velocity) are constant[30]. Therefore, by using Bernouli's principle, when the pressure is maximum, the velocity is minimum and visa versa, which means the pressure and velocity are 180° out of phase, inconsistent with sound waves. There is a version of Benouli's principal for unsteady flow[30] However, it has not been applied to playing a wind instrument as it requires knowing the time derivative of the velocity potential which is unknown and would require a solving the Euler equation.

Linguists have a similar issue in describing glottal vibrations and the accoustics of the vocal tract during speech. Recognizing Bernouli's principle does not apply, they make best efforts approximating it's effects, finding it modifies glottal vibration, but is not the driver[31]. They also note the acceleration of the air stream adds to the accoustic impedance of the system, thereby broadenning the frequency response.

This does not mean there is no pressure drop at the lip aperture, only that Bernouli's principle is inapplicable. The assumptions that go into Bernnouli's principle can be interpreted as the pressure distribution in the vocal tract and mouth are relatively uniform and time invarient, with only a changing pressure drop at the aperature. The uniform pressure distribution in the mouth is justified because in the frequency range brass players play, the wavelengths are longer than any dimension of the mouth. For example, at A7 (3520 cps), three octaves above A4 (440 cps) and the highest A on a standard eighty eight key piano, has a wavelength of almost 4 inches, larger than the dimensions of a typical oral cavity[31]. The vast majority of notes brass players produce are at lower frequencies and longer wavelengths. Therefore, we can keep the condition of a uniform pressure distribution in the mouth, but remove the constraint that the pressure in the mouth is time invarient. Therefore, in the mouth, the pressure distribution is uniform in space but varies with time. In other words, the mouth is an accoustic chamber, or a Helmholtz resonator. This model has fewer limiting assumptions than Bernouli's principle. In addition, the cup of the mouthpiece is another chamber with dimensions smaller than the wavelengths of playable notes and could also be modeled as a helmholtz resonator on the other side of the lips[3,5].

This is sensible from the point of view of brass players. Experienced brass players learn to drop their jaws and tongues for low frequency notes and raise them for high frequency notes[23]. They are constantly changing the volume of their mouths to match the pitch. Generally, brass players shape their mouths as if saying "Ohh" in the low register, moving to an "Ahh" sound to raise the pitch, and then to an "EEE" for the high register[11,14,23,32]. Without correction terms, the frequency of a Helmholz resonator is[31]:

$$F = \frac{c}{2\pi}\sqrt{\frac{S}{VL}}$$

Where $F$ is the frequency in cycles per second, $S$ is the area of the neck, $V$ is the volume of the chamber, and $L$ is the length of the neck. The range of Helmholtz resonance frequencies for the mouth and vocal tract can be estimated with differences for high and low ranges as shown in Figure 4. In the high range, brass players press their tongues close to the roofs of their mouths forming a narrow neck between the small volume behind their teeth and rest of the vocal tract. For low notes, brass players drop their jaws and the vocal tract above the lungs act as a neck while the lungs act as the Helmholtz volume. This would be much like head voice and chest voice in singing[24]. Head voice is the upper range where the vocalist feels a resonance in the their head and chest voice is in the lower range where the vocalist feels a resonance in their chest.

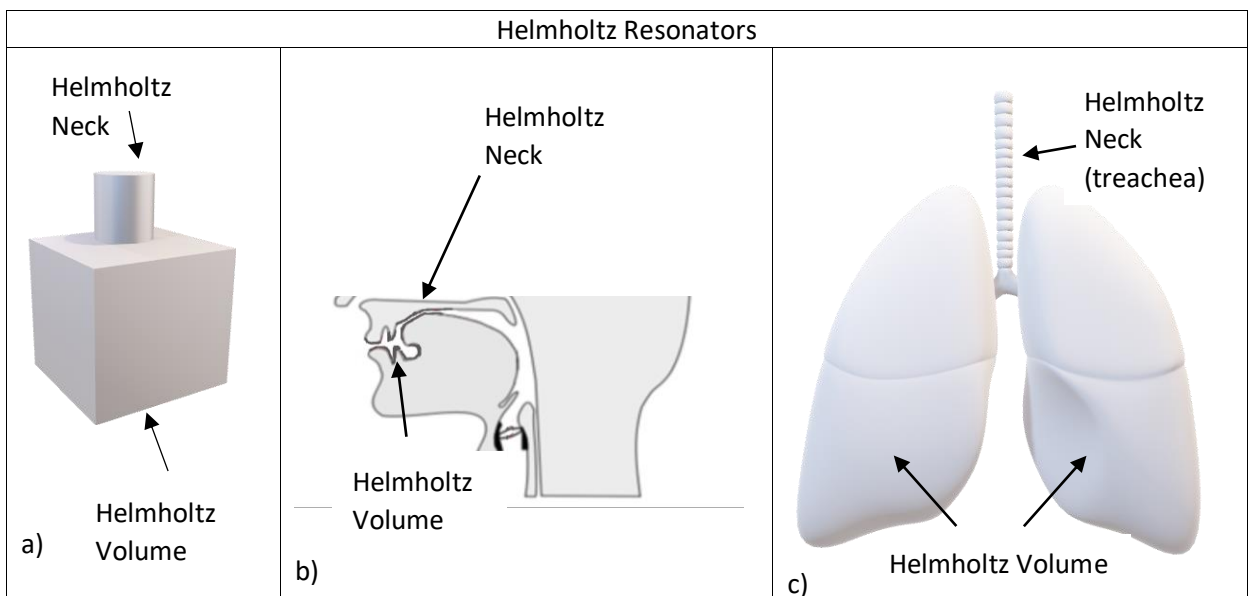

*Figure 4.* Different Helmholtz resonators. In a) a general description with a chamber and neck. In the high range as in b) the chamber is the volume between the teeth and the tongue while the neck is the volume between the tongue and the roof of the mouth. In the low range as in c) the chamber volume is the lungs while the neck is the trachea and the volume in the rest of the vocal tract.

There are combinations of mouth volumes and necks that have resonance frequencies in the the upper range of brass playing. The volume of a human mouth with the teeth closed is approximately 130 cc for females and 170 cc for males[31]. With the teeth open 1 cm it is approximately 150 cc for females and 190 for males[31]. The mouth volume may be made smaller by raising the tongue inside the mouth. For the neck (neck refers to the helmholz resonator, not the neck of a brass player), the area and length are needed. These can be approximated from the geometry of the narrow space between the tongue and the roof of the mouth. The width of the the space between the tongue and the roof of the mouth could be a centimeter or less and the distance between them a small fraction of a centimeter, then the area would be a fraction of a square centimeter. The length would be on the order of a centimeter or less. All these depend on how the tongue is pressed against the roof of the mouth. Figure 5 shows there are combinations of mouth volume, neck area, and neck length that would resonate at frequencies between 440 and 1760 cps.

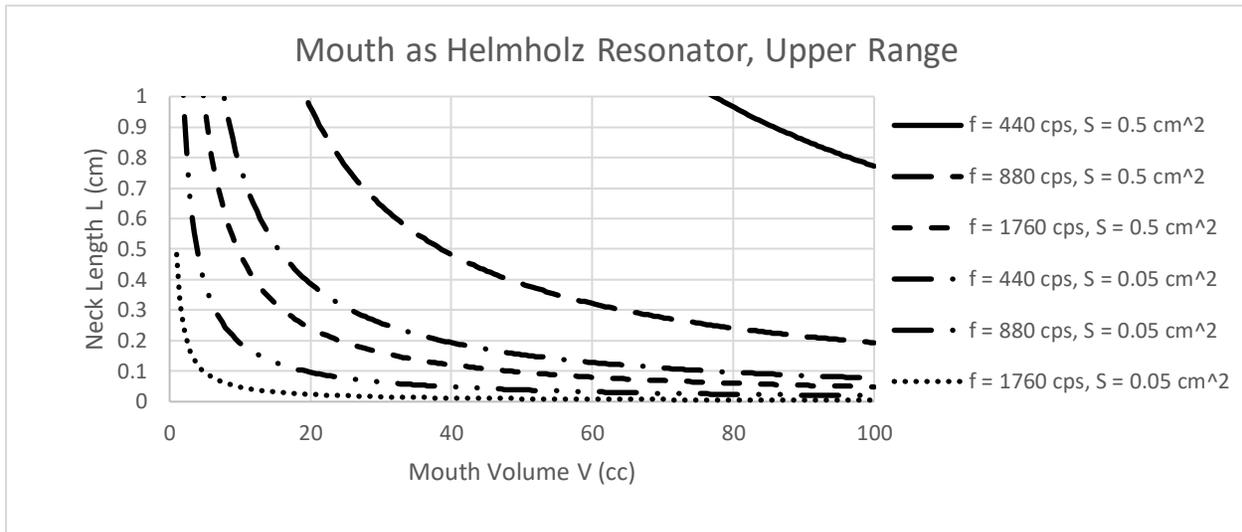

*Figure 5*. The mouth can be considered a Helmholtz resonator. For the upper range, the volume of the mouth is the volume of the Helmholtz chamber, while the space between the tongue and the roof of the mouth is the Helmholtz neck. Using a range of mouth volumes and reasonable choices for neck geometries, Helmholtz resonators with frequencies in the upper range of brass playing can be produced.

    For the lower range of brass playing, the mouth isn't a large enough resonating chamber for the low notes. One can get low frequencies if the lungs are used as the chamber volume. The residual lung capacity (the volume when a person has expelled all the air they can) is typically between 500 and 1000 $cm^3$, while the vital capacity (the amount that can be inhaled on top of the residual capacity) is typically between 3000 to 6000 $cm^3$ [31]. The treachea, mouth, nasal cavity, and sinuses are all volumes that can be considered part of the neck of the Helmholtz resonator. The area of the neck would be the area of the treachea, while the length of the neck would include the length of the treachea plus the effective lengths of the mouth, nasal cavity, and sinuses which would be their volumes divided by the area of the treachea. This would be much like finding the effective accoustic length of a mouthpiece in the low register: the volume divided by the area of the bore[5]. The typical area of a treachea is 2.5 $cm^2$, it's length is usually between 10 to 12 cm, the mouth volume can be up to 200 $cm^3$, and the nasal cavity sinuses can have a total volume of 110 $cm^3$ or larger[31]. This gives an effective length of the Helmholtz neck as up to 132 cm. Brass players can adjust this length by adjusting their mouth volume with their tongues and closing off or partially closing off the nasal passages with the velopharyngeal valve, as done during speech. This range of volumes can resonate at frequencies between 13 and 440 cps providing crossover with the high range Helmholtz resonator as shown in Figure 6.

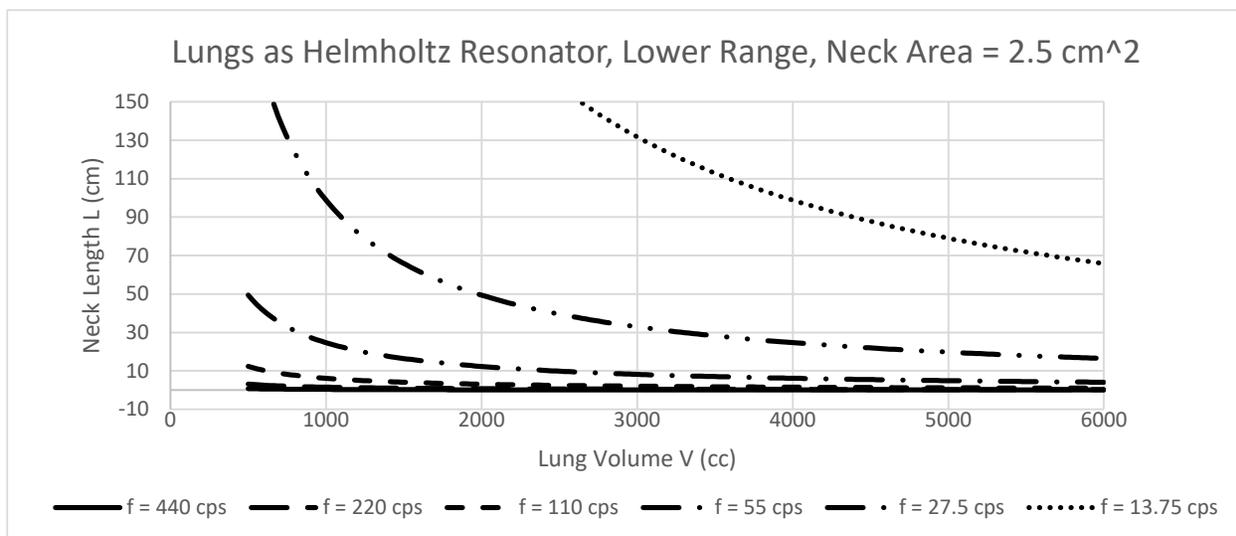

*Figure 6.* Using the lungs as the chamber of a Helmholz resonator, frequencies in the lower range of brass playing can be obtained using the vocal tract (trachea, mouth, nasal passage, and sinuses) as the neck.

To test whether the mouth and mouthpiece cup are resonating chambers, the pressures in the mouth, mouthpiece cup, mouthpiece backbore, and bell were measured simultaneously while playing. Figure 7 shows the mouthpiece with the transducers affixed and a separate transducer with a short tube for measuring pressure in the mouth. Figure 8 shows the pressure versus time traces for the author playing the first twelve harmonics of the trombone in first position (slide at its shortest). Going into the higher register increases the pressure inside the mouth. This is consistant with how brass players describe how it feels to play high notes. However, it is not simply an increase in static pressure with a small oscillation superimposed on top. The peak to trough value is increasing. In fact, since the gain was not changed , the signals in the mouth were getting clipped for the highest notes, even though the signal at the bell was decreasing.

Figure 9 shows the pressure traces for the author playing a Bb3 (233 cps or cycle time of 4.3 msec). One complete cycle is highlighted between zero crossings on the pressure trace of the cup. The pressure in the mouth is time varying, showing that Bernouli assumptions do not hold and the mouth is a resonating chamber. Pressures in the mouth and cup of the mouthpiece are 180 out of phase. That is sensible because they combine forces to drive the lips. If they were in phase, the lip would just compress and expand without opening. The cup and the backbore are also out of phase by 180. This shows the mouthpiece cup is a separate resonating chamber from the trombone: driving it and receiving feedback. Other descriptions of the the mouthpiece as a helmholtz resonator only use it to find the acoustic length of the mouthpiece using the popping frequency of the mouthpiece[5,6,33], not to see if it is driving the horn. It is treated as an extension of the horn but with varying diameter. If the mouthpiece were just that it would increase the acoustic length and one would expect the cup and the backbore to be in phase as they are only separated by` about 2.5 cm. However, the wavelength of the note being played is 147 cm. These same phase relationships were found to hold for all notes played from a Bb0 to an F5. The popping frequency of this mouthpiece was measured to be 550 cps. So this phase relationship held both above and below it. This doesn't mean the mouthpiece doesn't increase the acoustic length of the horn, but it is a separate accoustic element in series with it.

This description of the mouthpiece solves another problem in the brass literature. The horn is considered a closed tube at the mouthpiece end and open at the bell[4,5,33,34]. If that were the case, the horn would have resonances at only the odd multiples of of a fundamental frequency based on a wavelength four times the length of the horn[3,4,5,29]. But horns have resonances that are integer multiples of a fundamental based on the wavelength of twice the length of the horn. Those are what one would expect for a tube or horn open at both ends[3,4,5,29]. So an alternative explanation is that the mouthpiece is closed at the end with the lips, but it is a helmholtz resonator that drives the horn. In this description the horn is open at both ends because it is an air connection between the mouthpiece and the leadpipe of the instrument.

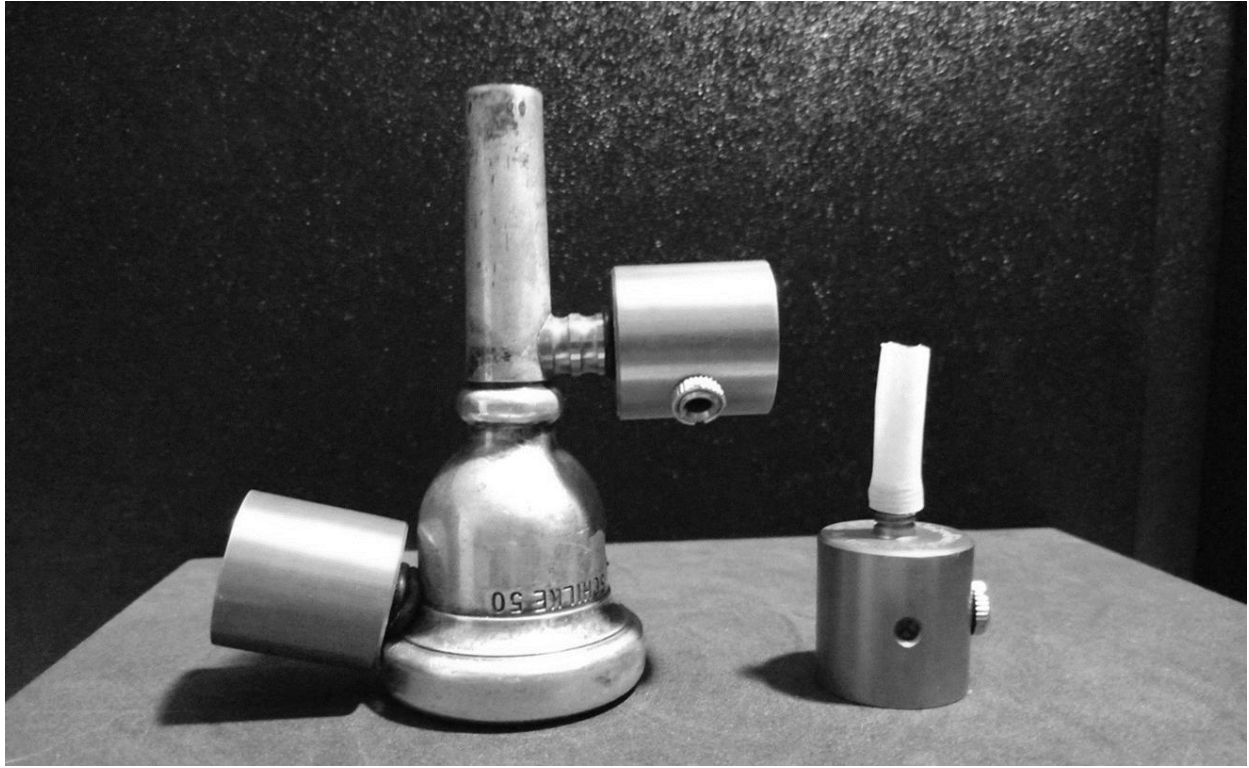

*Figure 7.* A Shilke 50 trombone mouthpiece with pressure transducers mounted in the cup and backbore. Also, a single transducer with a short plastic tube for insertion the mouth while playing. Transducers are from PiezoBarrel, Brisbane, Australia.

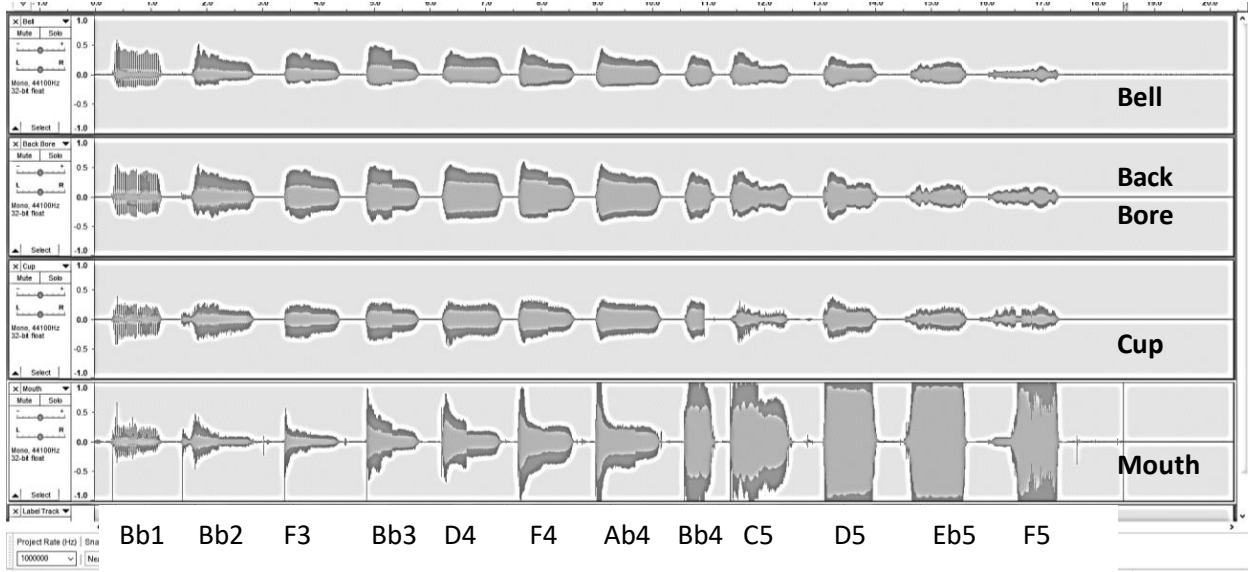

*Figure 8.* Pressure traced for the bell, backbore, cup, and mouth for the harmonic series in first position of the trombone, form the fundamental Bb1 to the 12[th] harmonic F6. Not only does the pressure in the mouth increase as the notes go up in frequency, but the peak to trough value increases. Some traces were shortened in time so they would all fit on one screen.

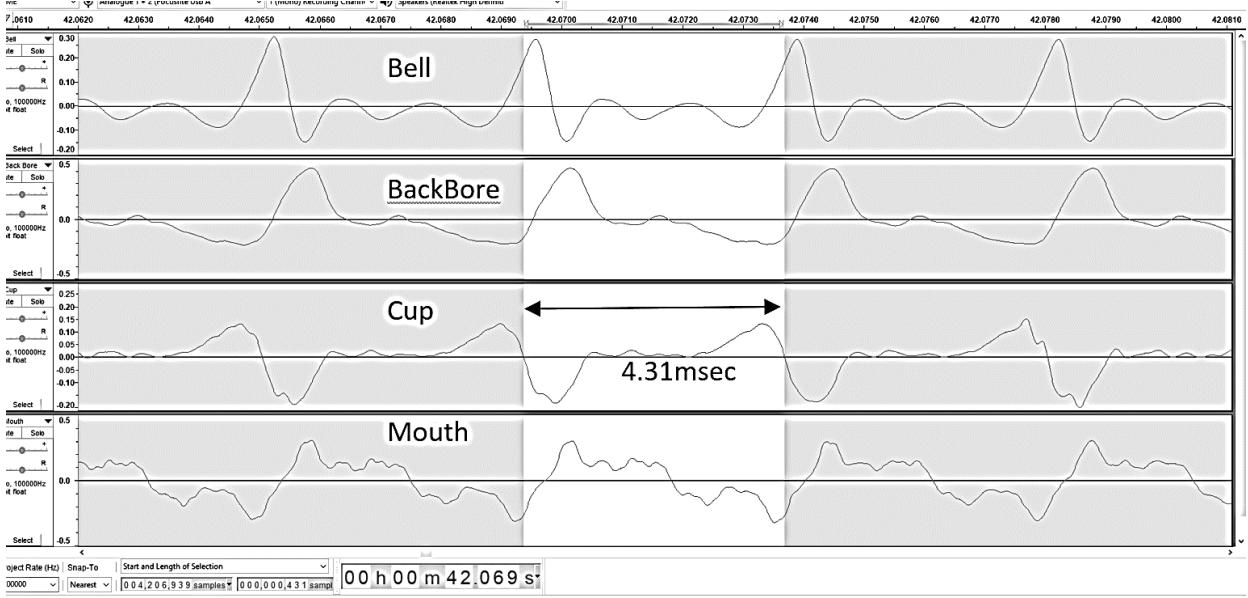

*Figure 9.* Synchronous pressure vs time signals for the mouth, cup, backbore, and bell while playing a Bb3 233 cps or 4.31 msec. Highlighted area is one cycle between zero crossings of the cup signal. The cup is very close to 180° out of phase with both the mouth and the backbore.

    A single note does not prove the mouth is driving the lips. The mouth pressure signal could be just following the lips. However, we can see the mouth leads when changing notes. Figure 10 is the time signature for a lip slur between a Bb3 (233 cps or cycle time of 4.3 msec) and D4 (294 cps or cycle time of 3.4 msec). In performing the slur, the author increases the air pressure and flow while raising his tongue and jaw. The first thing that happens is the pressure signal in the mouth increases even though the pressure signals down the line are decreasing. The frequency changes first in the mouth and then in the cup, backbore, and bell. The frequency in the mouth glisses to the higher note, while the signals

further down the line essentially stop and then restart at the higher frequency, as the new pulses make their way down the horn.

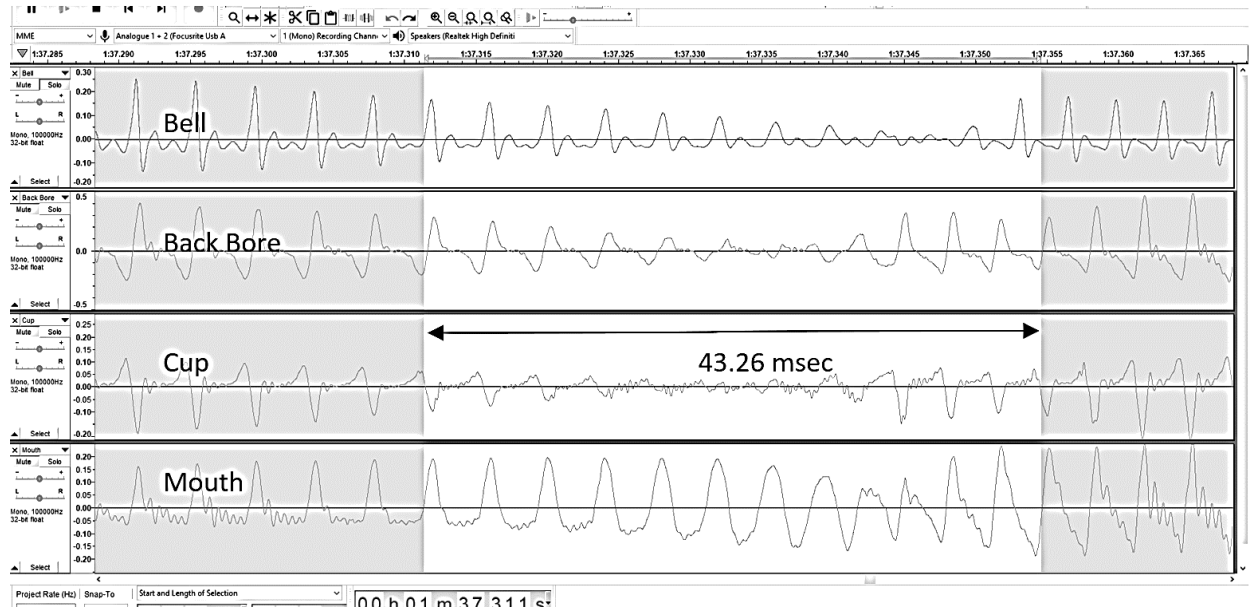

*Figure 10.* Synchronous pressure versus time signals for the mouth, cup, backbore, and bell while slurring between a Bb3 233 cps or 4.31 msec and a D4 at 294 cps or 3.4 msec.

Figure 11 shows the time trace of the fundamental of the trombone, Bb1, 58.3 cps. It has a cycle time of 17.32 msec which is the time it takes something traveling at the speed of sound to travel from the mouthpiece of the horn to the bell and back. The time trace shows a pulse approximately 3 msec wide. The pulse at the bell ocurs halfway between pulses at the backbore. The return pulse is in sync with the next pulse, so that the pulse in the cup is 180° of phase with the outgoing pulse from the mouth. This shows that the horn is acting like a transmission line with pulses traveling up and down it. This is consistant with a quality control technique developed for brass instruments to discover where imperfections are located by sending a pulse down the horn and measuring the time for any reflections. The distance to any imperfection is just half the time multiplied by the speed of sound[35].

The frequencies at which a pulse traveling the length of a tube and back would reinforce a new pulse is just:

$$f = \frac{nc}{2L}$$

Where $n$ is an integer, $c$ is the speed sound, and $L$ is the length of a tube. These are the same as the standing wave resonances for a tube of length L open at both ends[3,5,29]. At any other timings, the return pulses will hit the lip at the wrong time in its cycle making it difficult to play the note.

Input impedence studies in the frequency domain have shown that the fundamental frequency of a brass instrument should be significantly flat, and yet they aren't[3,5]. The usual explanation is based on "priviledged-tone" phenomena[4,33,34] where the upper harmonics dominate and the player excites the upper harmonics without exciting the fundamental. This work provides a simpler alternative explanation. Studies have shown the perceived pitch of a note is related to the periodicity of the signal[36]. By recognizing the time domain trace of the pressure is a pulse, the pitch of the fundamental remains in tune because pulses will have the time periodicity of twice the length of the horn divided by the speed of sound (really the group velocity of the pulse[4]).

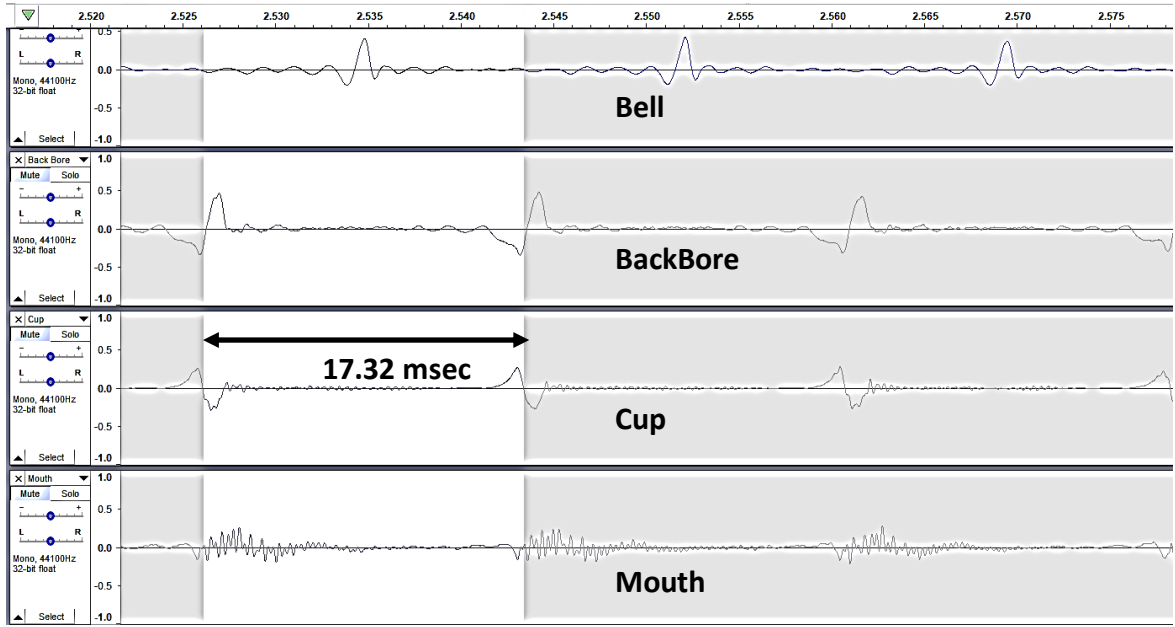

*Figure 11.* . Synchronous pressure versus time signals for the mouth, cup, backbore, and bell while playing the fundamental of the horn, Bb1, approximately 58.3 hz, which corresponds to a cycle time of 17.32 msec.  The signal at all measurement places are pulses, not sinusoids.

The 3 msec pulse width suggests there is an upper limit to the frequency if the pulse width stays the same.  A 3 msec cycle time is between a D4 (293 cps) and an F4 (349 cps) on the trombone harmonic series.  Figure 12 shows time traces for notes from D1 (36.7 cps) to Bb4 (466 cps).  All are harmonics in first position of the trombone except for F1 and F2 which were played in 1st position with the F attachment, and D1 which is in flat 4th position with the F attachment.  Over a 3 octave range from D1 to D4, the notes all have about the same pulse width and shape.  Only the spacings between pulses change.

This suggests an alternate way of viewing how brass instruments are played.  The traditional view is that the instrument has a series of standing wave modes and the lips are adjusted to the frequency of a particular mode and that note is played.  This data suggests the lips send a series of pulses down the horn in the lower 3 octaves and the frequency is increased by changing the spacing between pulses.  For the horn to resonate, the reflected pulse must reinforce the outgoing pulses at the lip.  These give the same modes as standing waves for a tube open at both ends.  In the higher range, the lips are propagating a traveling wave down the horn.  At the bell, some of the traveling wave is radiated and the rest is reflected back.  Again, there is reinforcement at lip at the same frequencies as the standing wave modes for a tube open at both ends.  In fact, it must be true because any standing wave may be decomposed into two traveling waves moving in opposite directions[28].

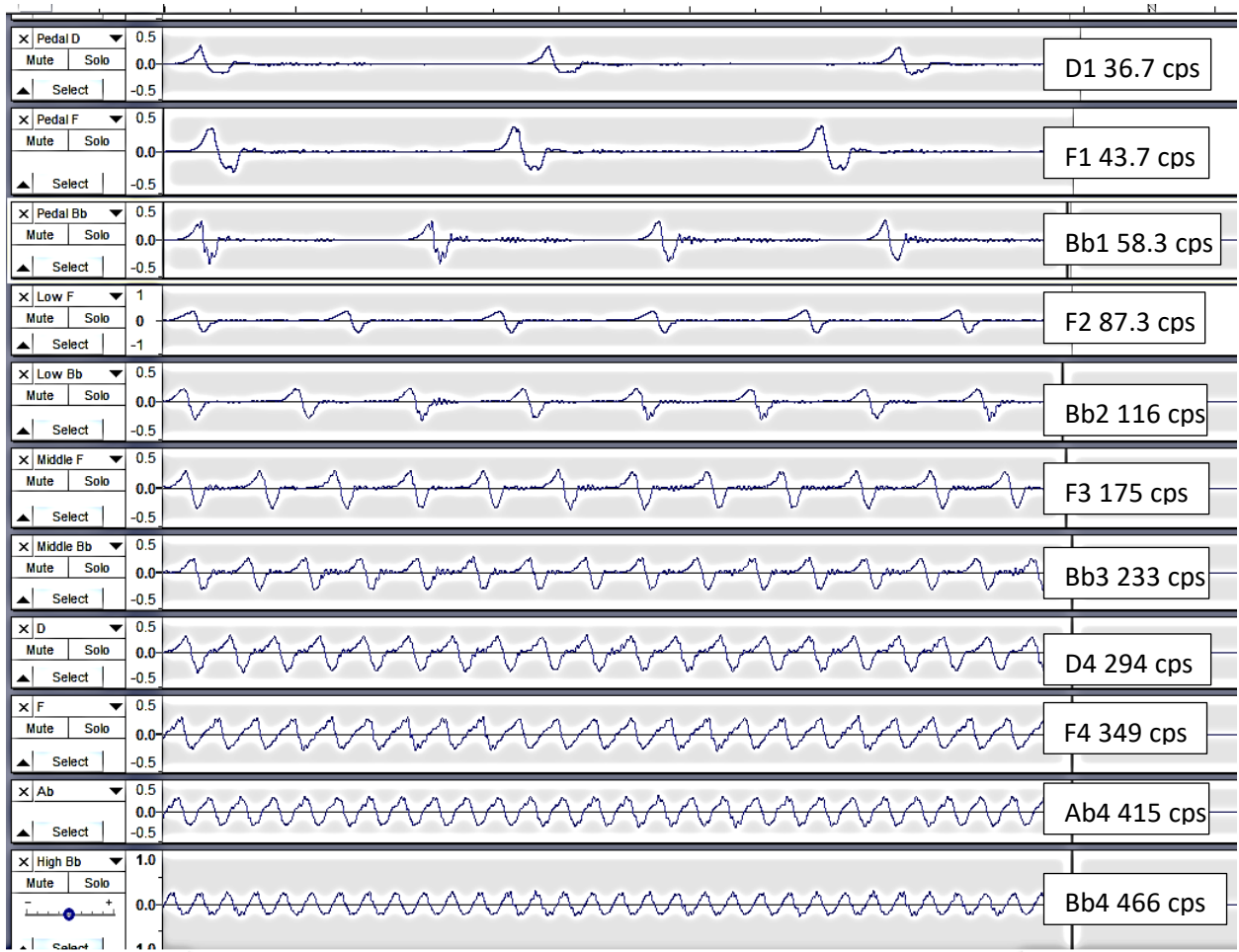

*Figure 12.* Time traces of pressure inside the cup of the mouthpiece for several different notes. Between D1 and D4, three octaves, the pulse width stays about the same, but the gap between pulses shortens. Above D4 the pulses get shorter, more like a continuous sinusoid.

There must be a change in what trombone players do around D3 because there is a shift from lessening the space between pulses to shortening the pulses for notes above D3. This is consistent with up or downstream playing which reduces the amount of lip that vibrated for the upper register. Earlier it was shown that while playing a Bb3 pushing the mouthpiece into the lips caused the frequency to double in accordance with the Farcas-Arban-Leno model. This was repeated except starting on a D3 and higher notes with the author's normal embouchure for each note. In these cases, pushing the mouthpiece into the lips caused the frequency to increase by about a half step or less: approximately 6% in frequency, not a doubling. First this means that once a player reduces the amount of lip that vibrates, pushing the mouthpiece into the lip no longer helps because the boundary is no longer at the rim of the mouthpiece but determined by the player. This suggests that by reducing the amount of lip that vibrates, upstream and downstream playing shorten the pressure pulse width. In other words, the pulse width is a function of the amount of lip that vibrates or the diameter of the vibrating portion of the lip.

In Figure 11, the pressure trace in the mouth has a high frequency component superimposed on the pulse. Its cycle time is approximately 0.32 msec. That corresponds to a frequency of 3136 cps or G7, the highest G on a standard piano. This means that while the lips are playing a Bb1, the mouth is ringing five and a half octaves above it! The wavelength of that frequency is 11 cm, or about the same as the distance between the teeth and the back of the mouth for an average male[31]. One can consider the high frequency as ringing inside a closed tube, the mouth.

Pulses can also explain the ability of brass players to play double pedal tones, or on the trombone a Bb0 at 29 cps. This note has a wavelength four times the length of a trombone or twice the wavelength corresponding to the fundamental frequency of the horn. It should not be playable from the viewpoint of a standing wave in the trombone. However, they can be explained as pulses. For a Bb1, a pulse goes out every 17.3 msec and that is the round-trip time for the pulse traveling the length of the trombone and back. If a pulse is sent out at half that rate, then the pulse returns halfway in the cycle when the lips are still open, and the return pulse is absorbed by the mouth. The pressure trace should show more noise in the mouth from absorbing the returning pulses. This can be seen in Figure 13. There is a smaller secondary pulse halfway between pulses representing the incomplete absorption of the pulse.

There is also more of the high frequency ringing in the mouth, and it bleeds into the cup. Figure 14 shows a comparison of the pressure traces in the cup of the mouthpiece for the pedal tone Bb1 and the double pedal tone Bb0. The double pedal tone has a broader pulse as well as more high frequency superimposed.

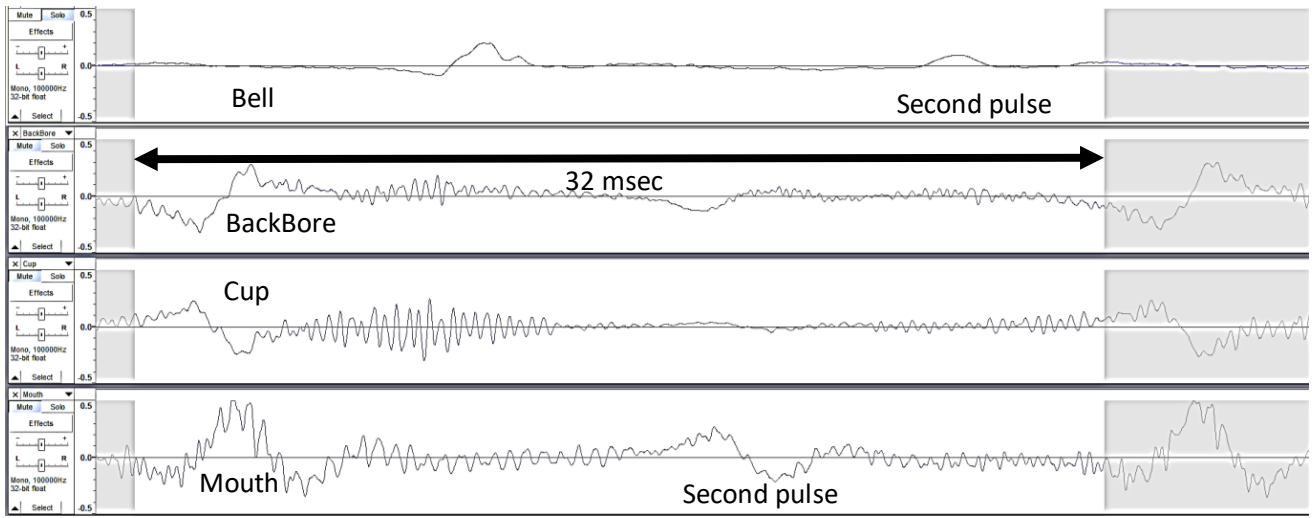

*Figure 13.* Time trace of pressure for a double pedal tone. There is a secondary pulse in the middle of the cycle not seen in the fundamental.

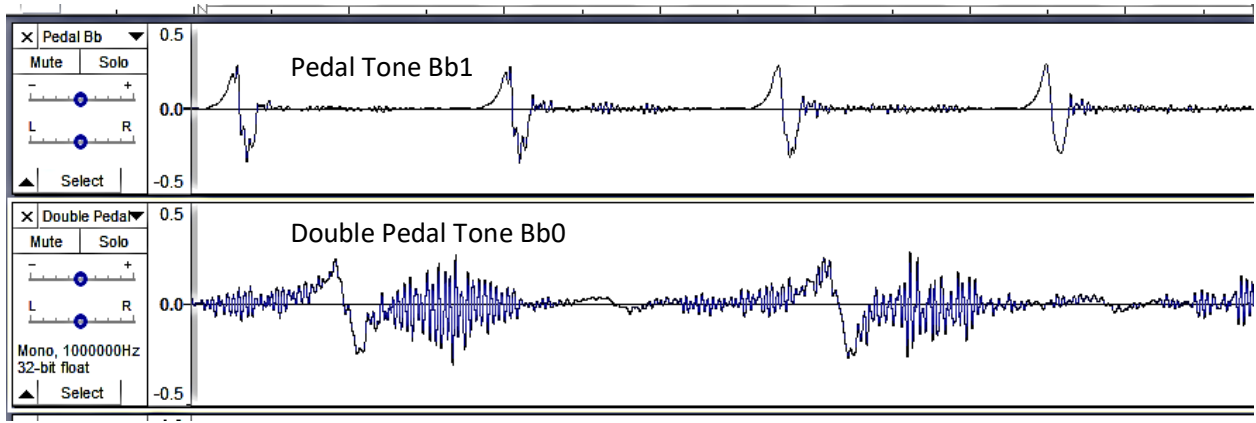

*Figure 14.* Comparison of the pressure traces in the mouthpiece cup for a pedal tone Bb1 and a double pedal tone Bb0. The pulse is broader and noisier for the double pedal tone.

## Summary/Conclusions

1. The change in modulus of the human lip between resting and fully contracted is not enough to explain the frequency range of brass players.
2. The lip can be modeled as a circular flat plate vibrating at its lowest mode. This fits the low end of brass player's frequency range, but not the high end.
3. Pushing the mouthpiece into the lips as inexperienced players often do, can increase the upper range by an octave, derived from flat plate theory and verified experimentally.
4. Experienced players use upstream and downstream playing to reduce the amount of lip that vibrates. This allows players to play higher frequencies without straining the primary lip muscle.
5. The lips are an overdamped oscillator.
6. The vocal tract is a resonating chamber and drives the lips to oscillate. Its resonance is modified by the player by changing their mouth volume and shape. This is done by raising and lowering the tongue and jaw.
7. The mouthpiece cup is a separate resonating chamber from the instrument, 180 out of phase with the instrument. It both drives the instrument and receives feedback from it, making the horn a tube that is open at both ends and driven by the mouthpiece at one.
8. The mouthpiece cup and the mouth are 180 out of phase which acts to efficiently drive the lips.
9. In the low range, notes are more accurately described as pulses and the horn acts as a transmission line, instead of standing wave modes.
10. Upstream and downstream playing are used to increase the frequency by shortening the pulses when they otherwise would overlap.
11. The pulse model can explain double pedal tones, which the standing wave model cannot.

## Acknowledgements

The author would like to thank Wayne Branco for his help in finding appropriate pickups, microphones, and general advice on trombone playing. Stephen Francis of Piezobarrel was incredibly helpful for all advice on mounting and using his pickups along with a wealth of technical information. Thanks to Dan Zwillinger for picking up inconsistencies, typos, and generally clarifying the math. Finally, the Nancy D. Kelly and S. Jay Keyser Foundation, for their generous financial support and encouragement for this work.

## Data Availability
All data is available from the corresponding author.